\documentclass[sigconf,final]{acmart}

\AtBeginDocument{%
  \providecommand\BibTeX{{%
    \normalfont B\kern-0.5em{\scshape i\kern-0.25em b}\kern-0.8em\TeX}}}

\setcopyright{acmcopyright}
\copyrightyear{2018}
\acmYear{2018}
\acmDOI{XXXXXXX.XXXXXXX}

\acmConference[SA '22]{SIGGRAPH Asia}{December 6--9,
  2022}{Daegu, South Korea}
\acmPrice{15.00}
\acmISBN{978-1-4503-XXXX-X/22/12}

\setcitestyle{square}

\usepackage[acronym]{glossaries}
\newacronym{ik}{IK}{Inverse Kinematics}
\newacronym{smpl}{SMPL}{SMPL}

\renewcommand{\vec}[1]{\mathbf{#1}}
\renewcommand{\Re}{\mathbb{R}}
\DeclareMathOperator{\smpl}{\textsc{SMPL}}
\DeclareMathOperator{\geodesic}{\textsc{GE}}
\DeclareMathOperator{\mpjpe}{\textsc{MPJPE}}
\DeclareMathOperator{\tr}{\textrm{tr}}
\DeclareMathOperator{\smplis}{\textsc{SMPL-SI}}
\DeclareMathOperator{\kernel}{\textsc{k}}

\usepackage{float}
\usepackage{placeins}

\begin{document}


\title{SMPL-IK: Learned Morphology-Aware Inverse Kinematics for AI Driven Artistic Workflows}
% \title{Learned SMPL Inverse Kinematics for AI Driven Artistic Workflows}

\author{Vikram Voleti}
\authornote{Authors contributed equally to this research.}
\email{vikram.voleti@umontreal.ca}
\affiliation{%
  \institution{Mila, University of Montreal}
  \country{}
}
\author{Boris N. Oreshkin}
\authornotemark[1]
\email{boris.oreshkin@gmail.com}
\affiliation{%
  \institution{Unity Technologies}
  \country{}
}
\author{Florent Bocquelet}
\authornotemark[1]
\email{florent.bocquelet@unity3d.com}
\affiliation{%
  \institution{Unity Technologies}
  \country{}
}

\author{F\'elix G. Harvey}
\affiliation{%
  \institution{Unity Technologies}
  \country{}
}

\author{Louis-Simon M\'enard}
\affiliation{%
  \institution{Unity Technologies}
  \country{}
}

\author{Christopher Pal}
\affiliation{%
  \institution{Mila, Polytechnique Montreal, CIFAR Chair, ServiceNow}
  \country{}
}

\renewcommand{\shortauthors}{Voleti, Oreshkin, Bocquelet, et al.}

\begin{abstract}
Inverse Kinematics (IK) systems are often rigid with respect to their input character, thus requiring user intervention to be adapted to new skeletons. In this paper we aim at creating a flexible, learned IK solver applicable to a wide variety of human morphologies. We extend a state-of-the-art machine learning IK solver to operate on the well known Skinned Multi-Person Linear model (SMPL). We call our model SMPL-IK, and show that when integrated into real-time 3D software, this extended system opens up opportunities for defining novel AI-assisted animation workflows. For example, pose authoring can be made more flexible with SMPL-IK by allowing users to modify gender and body shape while posing a character. Additionally, when chained with existing pose estimation algorithms, SMPL-IK accelerates posing by allowing users to bootstrap 3D scenes from 2D images while allowing for further editing. Finally, we propose a novel SMPL Shape Inversion mechanism (SMPL-SI) to map arbitrary humanoid characters to the SMPL space, allowing artists to leverage SMPL-IK on custom characters. In addition to qualitative demos showing proposed tools, we present quantitative SMPL-IK baselines on the H36M and AMASS datasets.

\end{abstract}

% TODO

% \ccsdesc[500]{Computing methodologies~Shape analysis}
% \ccsdesc[500]{Computing methodologies~Machine learning}


\begin{teaserfigure}
  \centering
  \includegraphics[width=\textwidth]{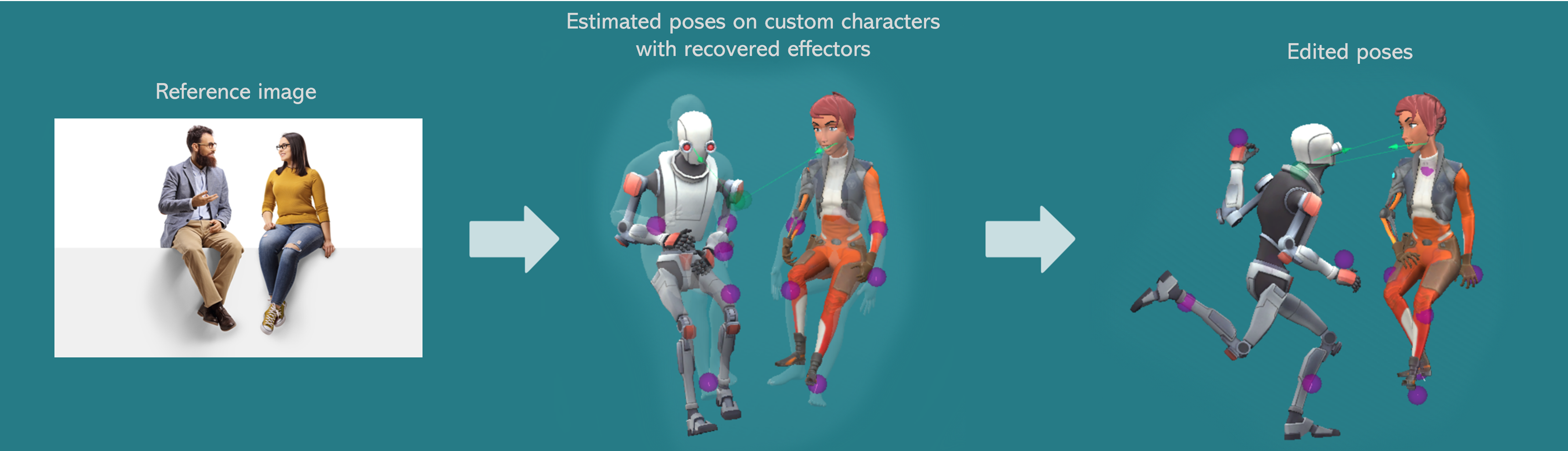}
  \caption{Our proposed morphology-aware inverse kinematics approach unlocks novel artistic workflows such as the one depicted above. Animator takes a photo of a multi-person scene or grabs one of the many pictures available on the web and uses it to initialize a 3D scene editable through a number of advanced animation tools. Our approach enables applying a multi-person 3D scene acquired from a RGB picture to custom user-defined characters and editing their respective 3D poses with the state-of-the-art machine learning inverse kinematics tool integrated in a real-time 3D development software.
%   The output of a monocular pose estimation algorithm is used to initialize the editing of a multi-person 3D scene. Our proposed SMPL-IK, SMPL-SI and effector recovery algorithms connect state-of-the-art AI pose authoring tools to user-defined custom characters and we use a computer vision backbone to refine the 3D scene acquired from a picture.
  }
  \Description{Two people seated.}
  \label{fig:teaser}
\end{teaserfigure}

\maketitle

\section{Introduction}

\gls{ik} is the problem of estimating 3D positions and rotations of body joints given some end-effector locations \cite{kawato1993supervised, aristidou2018inverse}. IK is an ill-posed nonlinear problem with multiple solutions. For example, given the 3D location of the right hand, what is a realistic human pose? It has been shown recently that machine learning IK model can be integrated with 3D content authoring user interface to produce a very effective pose authoring tool~\cite{oreshkin2021protores,bocquelet2022ai}. Using this tool, an animator provides a terse pose definition via a limited set of positional and angular constraints. The computer tool fills in the rest of the pose, minimizing pose authoring overhead.

The Skinned Multi-Person Linear (SMPL) model is a principled and popular way of jointly modelling human mesh, skeleton and pose~\cite{loper2015smpl}. It would seem natural to extend this model with inverse kinematics capabilities: making both human shape/mesh and pose editable using independent parameters. Additionally, many computer vision pose estimation algorithms naturally operate in the SMPL space making them natively compatible with a hypothetical SMPL IK model. This extension would open new content authoring opportunities. However, to date SMPL models have not been integrated with advanced machine learning IK tools, and this represents a clear research gap.

In our work we close this gap, exploring and solving two inverse problems in the context of the SMPL human mesh representation: SMPL-IK, an Inverse Kinematics model, and SMPL-SI, a Shape Inversion model. We show how these new components can be used to create new artistic workflows driven by AI algorithms. For example, we demonstrate the tool integrating SMPL-IK and SMPL-SI with an off-the-shelf image-to-pose model, initializing a multi-person 3D scene editable via flexible and easy-to-use user controls.

% First, we demonstrate the tool integrating SMPL-IK with an off-the-shelf image-to-pose model, and use that as a starting point for an editable SMPL character whose gender, shape and pose are editable via flexible and easy-to-use user controls. Furthermore, we show that by including the proposed SMPL-SI model in the workflow we add the additional flexibility of handling custom user supplied characters in the same universal SMPL space by finding an SMPL approximation of the user supplied character via SMPL-SI.

\section{Background}

\noindent\textbf{Inverse Kinematics} is a prominent problem in robotics and animation, which traditionally has been often solved by analytical or iterative optimization methods such as CCD~\citep{kenwright2012inverse} or FABRIK~\citep{aristidou2011fabrik} (please refer to~\citet{aristidou2018inverse} for a comprehensive review). Solving IK using machine learning techniques has consistently attracted attention~\cite{bocsi2011learning, d2001learning, de2008learning}, with more work focusing on neural networks based methods~\cite{duka2014neural,levine14,holden2016deep,
el2018comparative,bensadoun2022neural,mourot2022survey}. The current state-of-the-art IK neural network based approach is ProtoRes~\cite{oreshkin2021protores}. It takes a variable set of effector positions, rotations or look-at targets as inputs, and performs IK to reconstruct all joint locations and rotations. Its effectiveness in editing complex 3D character poses has recently been demonstrated within digital content creation software in a live demonstration~\cite{bocquelet2022ai}. Although ProtoRes is trained on MoCap data, it does not explicitly include any learnt body shape prior as it is trained on a fixed skeleton. In this work, we relax this limitation by integrating it with SMPL.

\noindent\textbf{SMPL} is a realistic 3D human body model parameterized by the body's shape and pose based on skinning and blend shapes \cite{loper2015smpl}. SMPL realistically represents a wide range of human body shapes controlled by shape parameters, as well as natural pose-dependent deformations controlled by pose parameters. 
%Hence, it can be used to efficiently animate realistic humans using existing animation tools.
Although there have been some extensions to the SMPL model such as SMPL+H~\cite{MANO:SIGGRAPHASIA:2017}, SMPL-X~\cite{pavlakos2019expressive},
STAR~\cite{STAR:2020}, 
etc., SMPL remains a standard model to represent realistic human body pose. SMPL is widely used for 3D pose estimation of humans in images and video~\cite{bogo2016keep,Luo_2020_ACCV,li2021hybrik,rajasegaran2021tracking,sun2021monocular}.

\noindent\textbf{Retargeting} is the task of transferring the pose of a source character to a target character with a different morphology (bone lengths) and possibly a different topology (number of joints, connectivity, etc.)~\cite{gleicher1998retargetting}. Retargeting is a ubiquitous task in animation, and procedural tools exist for retargeting between skeletons of different morphologies and topologies~\cite{unity2022retargeting}. 

% For example, retargeting is used to transfer the pose of a human captured using Motion Capture (MoCap) technology onto a user-supplied custom humanoid character.

\noindent\textbf{SMPL and IK} There is very little prior work that attempts to use the IK-enabled SMPL model for 3D character animation. \citet{bebko2021bmlSUP} pose SMPL characters in the Unity platform, but do not perform any IK. \citet{minimalIk} performs IK on SMPL parameters using standard optimization, but only in the full pose context, which has very limited applicability for artistic pose editing.  VPoser~\cite{pavlakos2019expressive} trains a Variational Auto-Encoder (VAE) to work as a prior on 3D human pose obtained from SMPL. This VAE is used as an iterative IK solver for a pose defined via keypoints. However, the VPoser architecture only works with relatively dense \emph{positional} inputs (no ability to handle sparse heterogeneous effector scenarios has been demonstrated). It also requires on-line L-BFGS optimization, making it too rigid and computationally expensive for pose authoring. There is a clear gap between SMPL and IK: existing IK models suitable for artistic pose editing do not support SMPL, and existing SMPL-based models have insufficient IK cababilities.
% as they were never intended for artistic pose editing.

% \FloatBarrier

\begin{figure*}[t] 
  \centering
  \includegraphics[width=\linewidth]{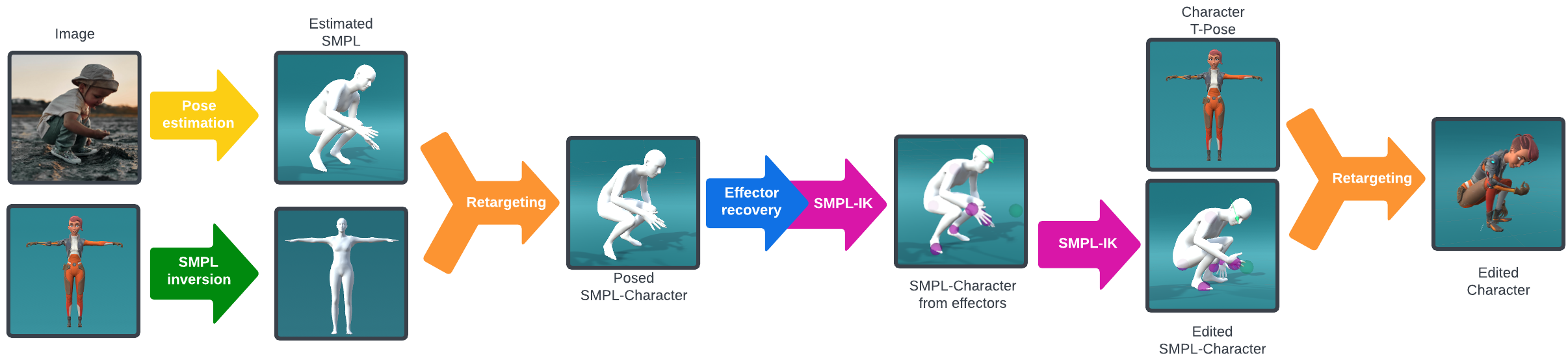}
  \caption{Pipeline for pose estimation and editing from a 2D image on a custom humanoid character.}
  \Description{Pipeline}
  \label{fig:pipeline}
\end{figure*}

% \FloatBarrier

\section{SMPL-IK} \label{sec:smpl-ik}
We propose SMPL-IK, a learned morphology-aware inverse kinematics module that accounts for SMPL shape and gender information to compute the full pose including the root joint position and 3D rotations of all SMPL joints based on a partially-defined pose specified by SMPL $\beta$-parameters (body shape), gender flag, and a few input effectors (positions, rotations, or look-at targets). SMPL-IK supports effector combinations of arbitrary number and type. SMPL-IK extends the learned inverse kinematics model ProtoRes~\cite{oreshkin2021protores}. ProtoRes only deals with a fixed morphology scenario in which an ML-based IK model is trained on a fixed skeleton. We remove this limitation by conditioning the ProtoRes computation on the SMPL $\beta$-parameters and gender
% , providing them as model inputs
(see Appendix~\ref{app:smpl_ik_neural_network} for technical details). This results in an IK model that can operate on the wide range of morphologies incorporated in the expansive dataset used to create the SMPL model itself.

There are multiple advantages of this extension, including the following. First, rich public datasets can be used to train a learned IK model, in our case we train on the large AMASS dataset~\cite{amass}. Second, an animator can now edit both the pose and the body shape of a flexible SMPL-based puppet using a state-of-the-art learned IK tool, which we demonstrate in Appendix~\ref{app:editing_both_shape_and_pose}. Third, training IK in SMPL space unlocks a seamless interface with off-the-shelf AI algorithms operating in a standardized SMPL space, such as computer vision pose estimation backbones.

\section{SMPL-SI} \label{sec:smpl_is}

SMPL-SI maps arbitrary humanoid skeletons onto their SMPL approximations by learning a mapping from skeleton features to the corresponding SMPL $\beta$-parameters (solving the inverse shape problem). Therefore, it can be used to map arbitrary user supplied skeletons in the SMPL representation and hence integrate SMPL-IK described above with custom user skeletons. Recall that the SMPL model implements the following forward equation: 
\begin{align} \label{eqn:smpl_forward_equation}
    \vec{p} = \smpl(\beta, \theta),
\end{align}
mapping shape parameters $\beta \in \Re^{10}$ and pose angles $\theta \in \Re^{22\times 3}$ into positions $\vec{p} \in \Re^{24\times 3}$ of SMPL joints. Datasets such as H36M contain multiple tuples $(\vec{p}_i, \beta_i, \theta_i)$. In principle, the pairs of H36M's skeleton features $\vec{f}_i$ extracted from  $(\vec{p}_i, \theta_i)$ and corresponding labels $\beta_i$, could be used for training a shape inversion model:
\begin{align}
    \hat\beta = \smplis(\vec{f}).
\end{align}
However, the H36M training set contains only 6 subjects, meaning that the entire dataset contains only 6 distinct vectors $\beta_i$, which is unlikely to be sufficient for learning any meaningful $\smplis$.

Taking this into account, we propose to train the $\smplis$ model as follows. We randomly sample 20k tuples $(\vec{p}_i, \beta_i)$ with $\widetilde\beta_i = [\epsilon_i, s_i]$, where $\epsilon_i \in \Re^{10}$ is a sample from uniform distribution $\mathcal{U}(-5, 5)$ and $s_i \in \Re$ is the scale sampled from $\mathcal{U}(0.2, 2)$. Scale $s_i$ accounts for the fact that the actual user supplied characters may have much smaller or much larger overall scale than the standard SMPL model. Furthermore, we use the $\smpl$ forward equation~\eqref{eqn:smpl_forward_equation} to compute joint positions $\widetilde{\vec{p}}_i$ corresponding to $\widetilde\beta_i$ and $\theta_i$ set to the T-pose. Finally, we compute skeleton features $\widetilde{\vec{f}}_i$ for each $\widetilde{\vec{p}}_i$ as distances between the following pairs of joints: (right hip, right knee), (right knee, right ankle), (head, right ankle), (head, right wrist), (right shoulder, right elbow), (right elbow, right wrist). Given the 20k samples from the $\smpl$ model and the features of the user skeleton $\vec{f}$, we implement the kernel density estimator for the shape parameters of the $\smpl$ model approximating the user supplied skeleton:
\begin{align} \label{smplsi_main_equation}
    \hat\beta = \sum_i \frac{\widetilde\beta_i w_i}{\sum_j w_j}, \quad w_i = \kernel((\vec{f} - \widetilde{\vec{f}}_i) / h).
\end{align}
Here $\kernel$ is the Gaussian kernel of width $h=0.02$. The theory behind this implementation is that in general, for each skeleton, characterized \emph{e.g.} by its bone lengths, there exist multiple equally plausible $\beta$'s (not surprisingly SMPL-SI is an ill-defined problem, like many other inverse problems). Therefore, a point solution of the inverse problem is likely to be degenerate. To resolve this, we formulate the general solution in probabilistic Bayesian terms, based on $p(\widetilde\beta, \vec{f})$, the joint generative distribution of skeleton shape and features. The corresponding posterior distribution of $\beta$ parameters given features gives rise to the following Bayesian $\beta$-estimator:
\begin{align}
    \hat\beta = \int \widetilde\beta p(\widetilde\beta | \vec{f}) d \widetilde\beta,
\end{align}
Note that $\hat\beta$ mixes a few likely values of $\widetilde\beta$ corresponding to posterior distribution modes. Decomposing $p(\widetilde\beta, \vec{f}) = p(\vec{f} | \widetilde\beta) p(\widetilde\beta)$ we get:
\begin{align} \label{eqn:beta_estimate_decomposition}
    \hat\beta = \int \frac{\widetilde\beta p(\vec{f} | \widetilde\beta) p(\widetilde\beta) d \widetilde\beta}{\int p(\vec{f} | \widetilde\beta) p(\widetilde\beta) d\widetilde\beta}. 
\end{align}
Since the joint distribution $p(\widetilde\beta, \vec{f})$ is unknown, we approximate it using a combination of kernel density estimation and Monte-Carlo sampling. Assuming conservative uniform prior for $p(\widetilde\beta)$, we sample $\beta$ as described above and we use a kernel density estimator $p(\vec{f} | \widetilde\beta) \approx \frac{1}{hN} \sum_i \kernel(\frac{\vec{f} - \widetilde{\vec{f}}_i}{h})$. Using this in~\eqref{eqn:beta_estimate_decomposition} together with Monte-Carlo sampling from $p(\widetilde\beta)$, results in~\eqref{smplsi_main_equation}.

\section{Proposed AI driven Workflow} \label{sec:proposed_ai_driven_flow}

Figure~\ref{fig:teaser} presents a high-level summary of the proposed artistic workflow for 3D scene authoring from an image, while Fig.~\ref{fig:pipeline} provides the detailed overview of how it is implemented for a user-defined humanoid character. Appendices~\ref{app:editing_both_shape_and_pose},~\ref{app:pose_authoring} and~\ref{app:labeling_tool_demo} depict simpler workflows for authoring SMPL poses, image labeling in the SMPL space and authoring poses on custom characters from scratch. These were implemented in the 3D real-time Unity engine for validation. These workflows leverage SMPL-IK and SMPL-SI building blocks as well as some others described in the rest of this section.

% \subsection{3D Scene From Image} \label{ROMP}

\textbf{3D Scene From Image}. We propose to process a monocular RGB image to initialize an editable 3D scene as shown in Fig.~\ref{fig:teaser}. A few methods exist for pose estimation from RGB inputs, most recent of which include ROMP~\cite{sun2021monocular} and HybrIK~\cite{li2021hybrik}. In our approach, we use a pre-trained ROMP model that predicts shape, 3D joint rotations and 3D root joint location for each human instance in the image. The outputs of pose estimation can be directly used to edit the estimated 3D SMPL mesh using SMPL-IK, leading to advanced 3D labelling tools that can be used to refine pose estimation and augmented reality datasets, as described in Appendix~\ref{app:labeling_tool_demo}. Alternatively, pose estimation results can be retargeted to user-supplied 3D characters. In which case, the 3D scene with retargeted characters is further edited through the combination of SMPL-IK and SMPL-SI as explained below.

\textbf{Custom Characters}. Pose estimation algorithms output pose in the standardized SMPL space, whereas users may wish to repurpose the pose towards their own custom character. We use SMPL-SI to find the best approximation of the custom character by estimating its corresponding SMPL $\beta$ parameters from the custom skeleton features (e.g. certain bone lengths). The SMPL character created using SMPL-SI provides a good approximation of the user character hence providing for its smooth integration with SMPL-IK, operating in the standard SMPL space.

\textbf{Retargeting}. In Figure~\ref{fig:pipeline}, procedural retargeting first retargets the initial pose estimation result onto the SMPL approximation of the user-defined character obtained via SMPL-SI. Second, it retargets the pose edited by the animator with SMPL-IK back on the user character. On both occasions, SMPL-SI makes the job of procedural retargeting easier. First, it aligns the topology of user character with the SMPL space. Second, the SMPL character derived via SMPL-SI is a close approximation of the user character, simplifying the transfer of the pose edited with SMPL-IK back onto the user character.

% We propose to handle the retargeting task in two steps: (i) ML-based topology alignment, followed by (ii) traditional procedural retargeting in SMPL space. We prTopology alignment is done via an ML-based technique called SMPL Inversion proposed in this paper (described below). SMPL Inversion is used to find an SMPL mesh that best fits the user-supplied custom character. Procedural morphology retargeting is used for retargeting between different SMPL skeletons via standard Unity tool~\cite{}???.

\textbf{Effector Recovery}. Pose estimation outputs a full pose (24 3D joint angles and 3D root joint location) of each human in the scene. The pose editing process constrained by this information would be very tedious. SMPL-IK makes pose authoring efficient using very sparse constraints (e.g. using 5-6 effectors). Therefore, we propose to extract only a few effectors to create an editable initial pose. We call this \emph{Effector Recovery}, which proceeds starting from an empty set of effectors, given the full pose provided by the computer vision backbone, in an iterative greedy fashion. Out of the remaining effectors, we add one at a time, run a new candidate effector configuration through SMPL-IK, and obtain the pose reconstructed from this configuration. We then choose a new effector configuration by retaining the candidate effector set that minimizes the L2 joint reconstruction error in the character space. We repeat this process until either the maximum number of allowed effectors is reached, or the reconstruction error falls below a fixed threshold. We find this greedy algorithm very effective in producing a minimalistic set of effectors most useful in retaining the initial pose, which is shown in supplementary video discussed in Appendix~\ref{app:effector_recovery}.

\textbf{Pose Editing}. Pose editing relies on the Unity UX integration of SMPL-IK similar to one of ProtoRes and augmented with the SMPL shape editing controls as well as pose estimation, SMPL-SI, retargeting and effector recovery integrations. Editing happens directly in the user character space following the WYSIWYG paradigm. A full pipeline demo is presented in Appendix~\ref{app:full_pipeline_demo}.

\section{Empirical Results}

\begin{table}[t] 
    \centering
    \caption{SMPL-IK baseline for the new benchmark following the randomized effector scheme~\cite{oreshkin2021protores} implemented on AMASS and H36M datasets, based on MPJPE (Mean Per Joint Position Error), PA-MPJPE (Procrustes-Aligned MPJPE), and GE (Geodesic Error) metrics.}
    \label{table:protorez_vs_baselines}
    \begin{tabular}{ccc|ccc} 
        \toprule
        \multicolumn{3}{c|}{AMASS} & \multicolumn{3}{c}{H36M} \\
        \cmidrule(lr){1-3} \cmidrule(lr){4-6} 
        MPJPE & PA-MPJPE  & GE & MPJPE & PA-MPJPE  & GE \\
        \hline
        59.3 & 52.5 & 0.1602  & 65.8 & 57.9 & 0.224  \\ 
        \bottomrule
    \end{tabular}
\end{table}

In \autoref{table:protorez_vs_baselines}, we report pose reconstruction errors of our SMPL-IK approach for two datasets : AMASS~\cite{amass} and Human3.6M~\cite{h36m_pami}
%The metrics used to evaluate SMPL-IK are Mean Per Joint Position Error (MPJPE), Procrustes-Aligned MPJPE (PA-MPJPE), and Geodesic Error (GE, see e.g.~\citet{salehi2018real} for the definition). 
(see Appendix~\ref{app:splik_training_details} for more details).

\section{Limitations}

SMPL-IK and SMPL-SI are most effective when dealing with realistic human shapes and poses, because they are trained on the SMPL model and realistic 3D pose data from the AMASS dataset~\cite{amass}. Obviously, they perform worse when dealing with unrealistic and disproportionate human body types, such as those of certain cartoon characters. 
%This also implies that effector recovery is limited when retargeting 3D characters that are not realistic. 
SMPL-SI relies on a set of joints to compute user character features. These joints are present in most characters, but without them its operation is not viable. 

% \begin{acks}
% \end{acks}

\bibliographystyle{ACM-Reference-Format}
\bibliography{ref}

\clearpage

\appendix

All demo videos are available here: \url{https://drive.google.com/drive/u/1/folders/1bHwoZjAX9njFCGszzLpUtOFGXxs0sWKW}.

\section{Editing Both Shape and Pose Demo} \label{app:editing_both_shape_and_pose}
 \begin{figure}[h]
  \centering
  \includegraphics[width=\linewidth]{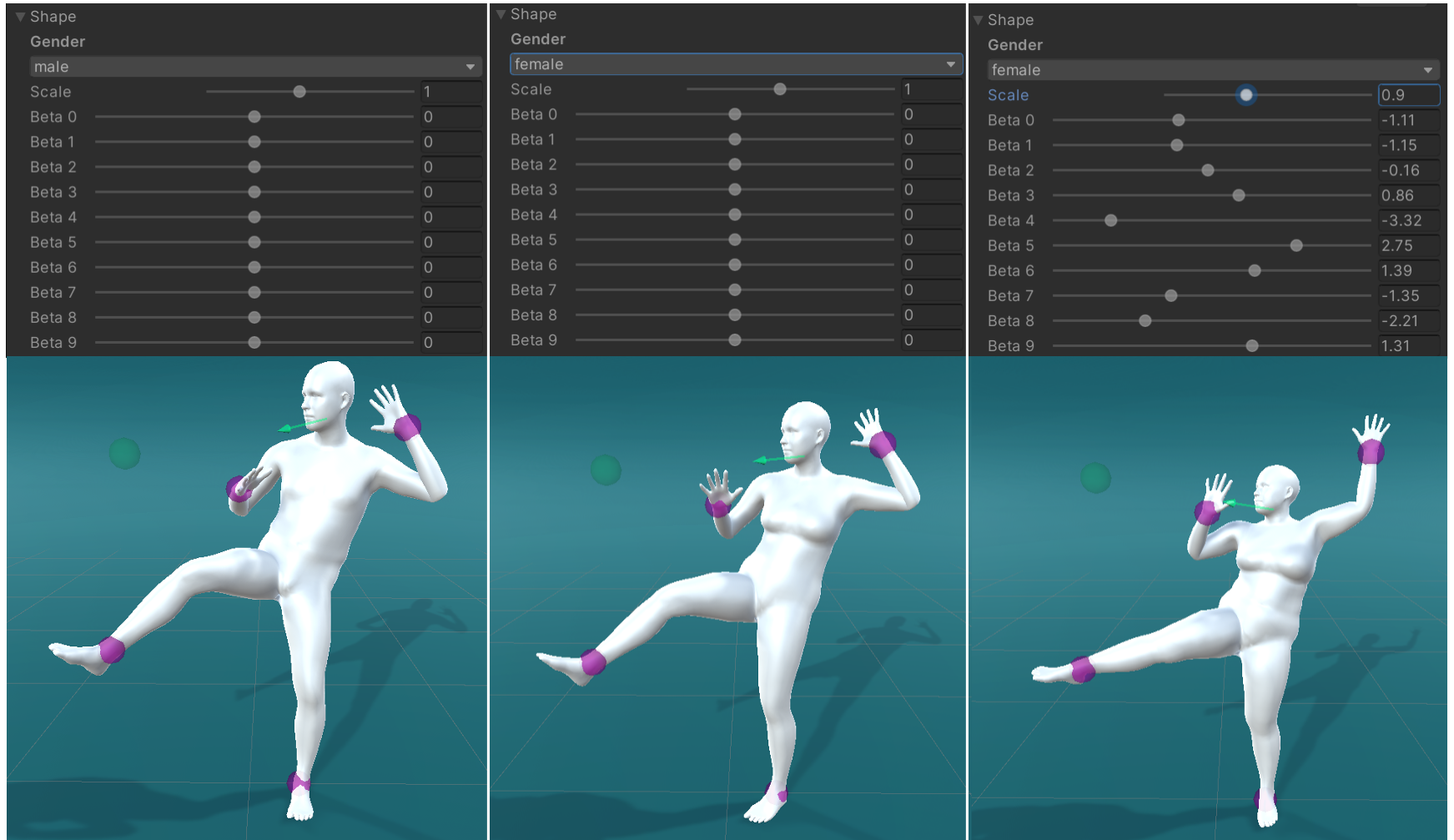}
  \caption{Morphology-aware learned IK. Left: posing the average male SMPL character. Center: result of modifying only the SMPL gender parameter. Right: result of additionally modifying the SMPL $\beta$ shape parameters.}
  \Description{Pipeline for pose authoring with a custom humanoid character via SMPL-IK and SMPL-SI.}
  \label{fig:shape_and_demo_posing}
\end{figure}
This is shown in the supplementary video \url{Demo\_Pose\_and\_Shape\_Editing.mp4}. Compared to ProtoRes, SMPL-IK adds the additional flexibility of editing body shape together with pose. Figure~\ref{fig:shape_and_demo_posing} demonstrates the user interface of shape editing, including the gender setting and the controls for the 10 SMPL $\beta$ parameters. In addition, the demo video shows how pose and shape of the SMPL character can be edited simultaneously. In this video, we demonstrate the benefit of SMPL-IK in pose authoring. First, we show how different effectors can be successfully manipulated using SMPL-IK leading to different realistic poses of the same body. Then, we show how changing body type, described by gender and scale, leads to different realistic versions of the same pose for different bodies. Finally, we show fine-grained modification of the body type by manipulating the SMPL shape parameters of the body. At every step, the corresponding pose is estimated using our SMPL-IK approach.

\section{Labeling tool demo} \label{app:labeling_tool_demo}
\begin{figure}[h]
  \centering
  \includegraphics[width=\columnwidth]{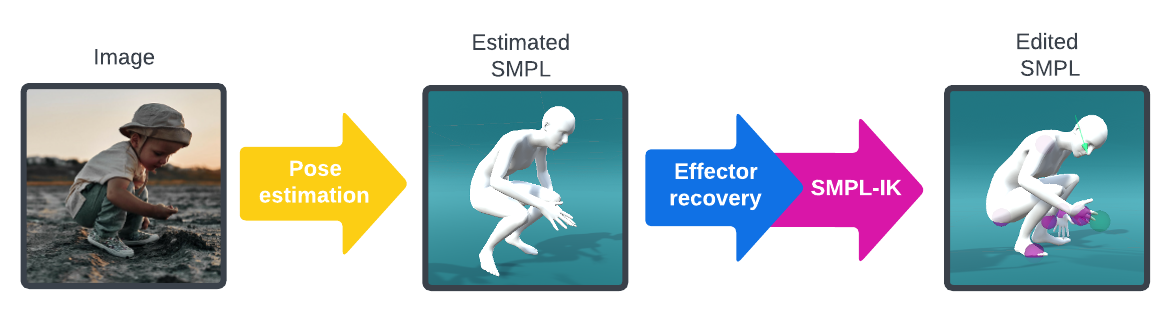}
  \caption{Pipeline for 2D image labeling with accurate 3D pose.}
  \Description{Pipeline for 2D image labeling with accurate 3D pose.}
  \label{fig:pipeline_labeling}
\end{figure}

This is shown in the supplementary video \url{Demo\_Labeling\_Tool.mp4}. One of the drawbacks of the current pose estimation datasets based on real data is that only 3D or 2D positions of joints are actually labeled. However, it was shown that rotations are very important for representing a believable naturally looking pose~\cite{oreshkin2021protores}. SMPL-IK can be used as a labeling tool to add the missing 3D-rotation information to existing datasets, elevating them to the next level with minimal human effort. Given an image of a human, our SMPL-IK approach (combined with an off-the-shelf image-to-pose estimator) provides an editable 3D SMPL model in a pose close to the one in the image  (see Fig.~\ref{fig:pipeline_labeling}). The labeling tool based on SMPL-IK and its integration with Unity can be used to correct the joint rotations and specify the correct lookat (head/eyes direction) that is most often estimated incorrectly by the current SOTA pose estimation algorithms due to the absence of this information in the current pose estimation datasets.

 \section{Pose authoring on a custom character} \label{app:pose_authoring}
\begin{figure}[h]
  \centering
  \includegraphics[width=\linewidth]{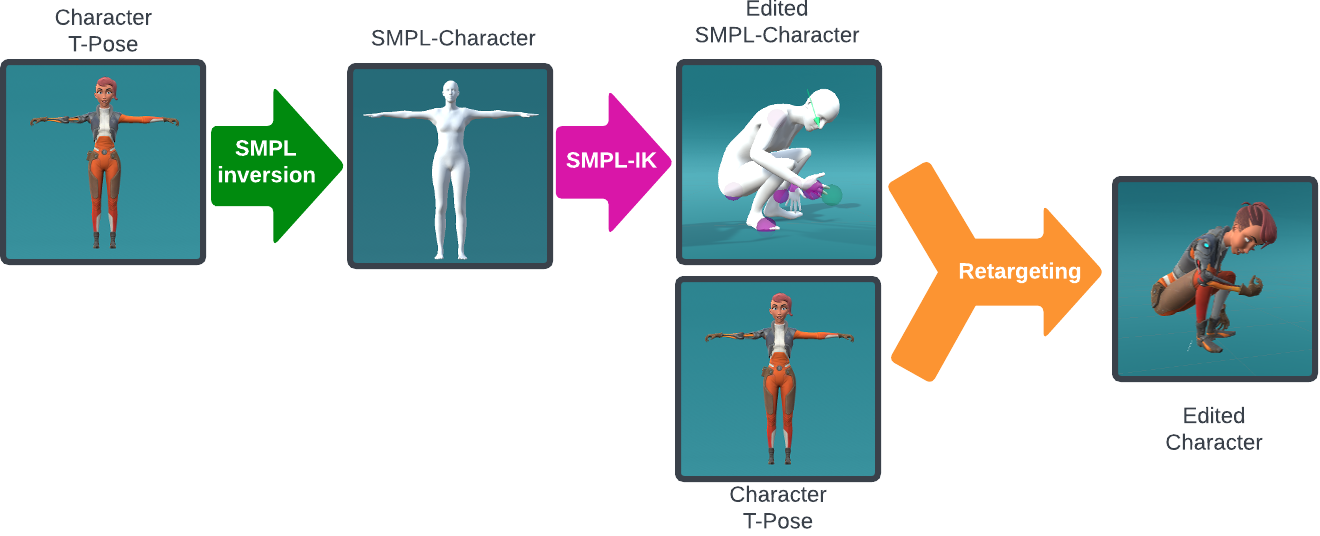}
  \caption{Pipeline for pose authoring with a custom humanoid character via SMPL-IK and SMPL-SI.}
  \Description{Pipeline for pose authoring with a custom humanoid character via SMPL-IK and SMPL-SI.}
  \label{fig:pipeline_custom_posing}
\end{figure}

Figure~\ref{fig:pipeline_custom_posing} depicts the simplified workflow that is used for authoring a pose for the custom user defined character using a combination of SMPL-IK and SMPL-SI. Supplementary videos \url{Demo_Authoring_Pose_Child.mp4},  \url{Demo_Authoring_Pose_Child.mp4}, \url{Demo_Authoring_Pose_Female.mp4}, \url{Demo_Authoring_Pose_Male.mp4}, \url{Demo_Authoring_Pose_Strong.mp4} show how SMPL-SI can be used to manipulate 4 custom characters (child, female, male and strong male) with different proportions and morphologies.

\section{Effector Recovery} \label{app:effector_recovery}

Supplementary video \url{Demo_Effector_Recovery.mp4} demonstrates the effector recovery mechanism in action. It shows the effect of changing the maximum number of effectors hyperparameter as well as the effect of changing number of recovered effectors on the initial pose extracted from image. It is clear that a relatively small number of effectors are sufficient to recover a good initialization for the editable pose.

\section{Full Pipeline Demo} \label{app:full_pipeline_demo}

\subsection{Demo\_Crouch\_FineTuning.mp4}

In this video, we show how to edit a pose in Unity using our approach. First, given a user-provided 3D character, SMPL-SI is used to estimate the SMPL body shape parameters that best fit the character. This SMPL-Character is shown in the video transparently along with the 3D character, and is also shown in the second image in \autoref{app:pose_authoring}. Then, given an image of a human in a pose, such as the crouched baby in \autoref{app:labeling_tool_demo}, an off-the-shelf image-to-pose estimator is used to obtain its SMPL pose parameters. Then, the SMPL-Character is retargeted onto the estimated pose. Next, for further editing of the character from the new pose, Effector Recovery is performed to recover the best effectors that describe that pose for that character. The effectors are shown in purple. These effectors can now be used to edit the pose as the user wishes. Optionally, more effectors could be activated for further fine-tuned editing, including both positional and rotational effectors.

\subsection{Demo\_Sitting\_Editing.mp4}

In this video, we demonstrate the case shown in the teaser image, with two humanoid 3D characters. As image of two people sitting is loaded, the poses of the two people are estimated using an off-the-shelf image-to-pose model, and the two 3D characters are retargeted to these estimated poses. Further, the pose of the 3D characters are then edited by manipulating the effectors. The video shows the various effectors and the effect of manipulating them. Every manipulation uses our SMPL-IK approach to estimate the realistic pose of that character.

\section{SMPL-IK Details} \label{app:smpl_ik_details}

\subsection{SMPL-IK Neural Network Diagram} \label{app:smpl_ik_neural_network}

\begin{figure}[h]
  \centering
  \includegraphics[width=\columnwidth]{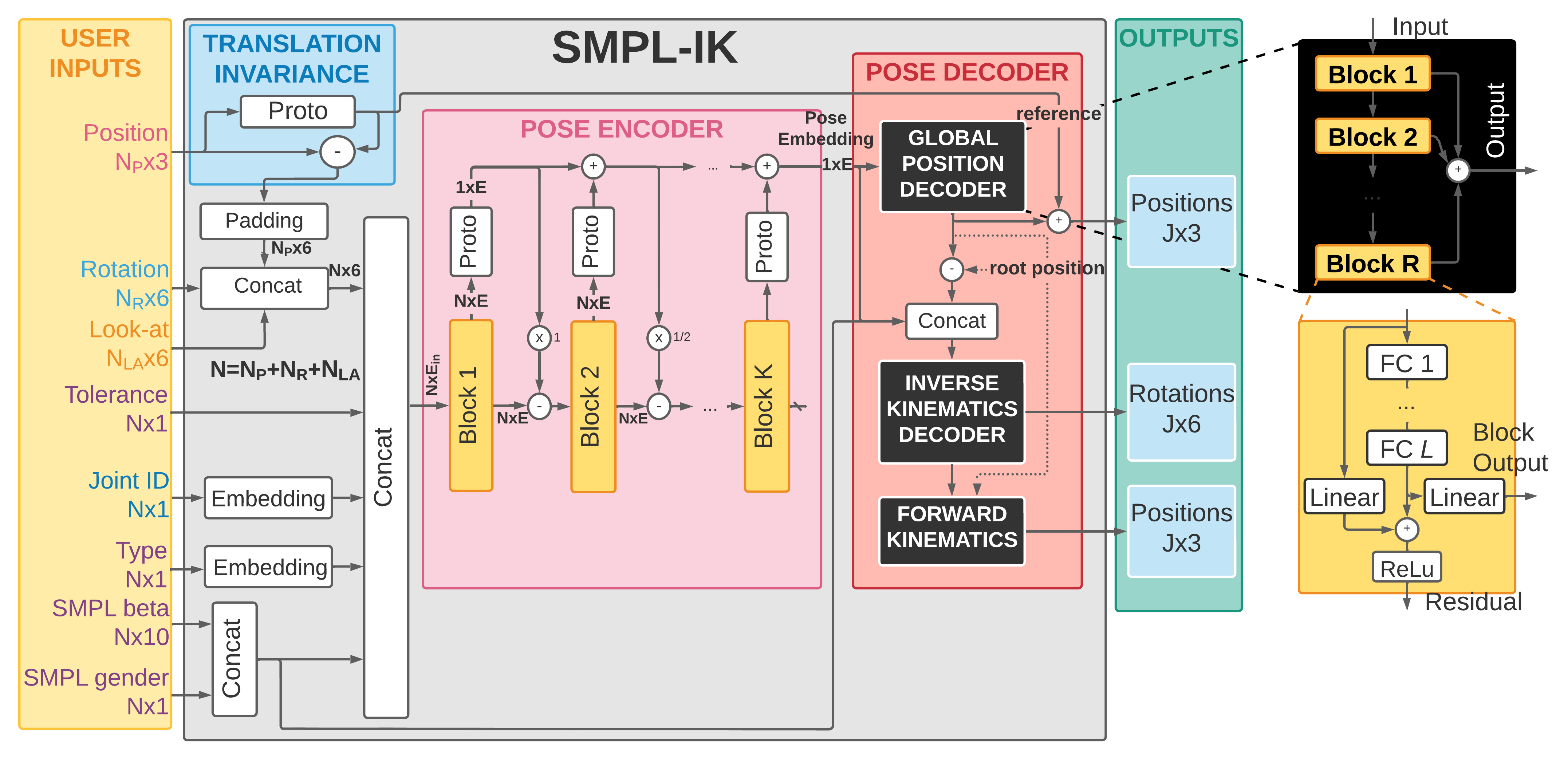}
  \caption{SMPL-IK neural network diagram. Note input conditioning on $\beta$ and gender inputs.}
  \label{fig:smpl_ik_neural_network}
\end{figure}

\subsection{SMPL-IK Training and Evaluation Details} \label{app:splik_training_details}

The training process overall follows that of ProtoRes~\cite{oreshkin2021protores}. We found that naively training on AMASS works well, while for H36M, simply training on its 6 subjects results in the lack of generalization in the $\beta$ subspace of inputs. To overcome this, we used the following $\beta$ augmentation strategy. For each sample drawn from the H36M dataset we added white Gaussian noise with unit variance and recalculated joint positions based on the augmented value of $\beta$ and the pose $\theta$ from the dataset. The model was trained on the augmented H36M dataset. We found that overall, the qualitative model quality was better when it was trained on the AMASS dataset, although the quality of the H36M model was also acceptable. 

To measure quantitative generalization results, we used H36M train and test splits derived in ROMP~\cite{sun2021monocular}, which in tern follow H36M Protocol 2 (subjects S1, S5, S6, S7, S8 for training and S9, S11 for test, plus 1:10 subsampling of the training set) and AMASS train/validation/test splits from~\cite{amass} (valdation datasets: HumanEva, MPI\_HDM05, SFU, MPI\_mosh; test datasets: Transitions\_mocap, SSM\_synced; training datasets: everything else).

To quantify SMPL-IK, we choose to use evaluation metrics commonly used in the context of H36M and AMASS datasets, MPJPE and PA-MPJPE plus the geodesic rotation error, which was shown to be important in quantifying the quality of realistic poses in~\cite{oreshkin2021protores}. The metrics are defined as follows. 

GE, geodesic error, between a rotation matrix $\vec{R}$ and its prediction $\widehat{\vec{R}}$,~\citet{salehi2018real}:
\begin{align} \label{eqn:geodesic_loss}
\geodesic(\vec{R}, \widehat{\vec{R}}) = \arccos\left[(\tr(\widehat{\vec{R}}^T \vec{R}) - 1) / 2  \right].
\end{align}
MPJPE, mean per joint position error, is computed by flattening all poses and joints in the batch into the leading dimension resulting in the ground truth tensor $\vec{p} \in \Re^{N \times 3}$ and its prediction $\widehat{\vec{p}}$:
\begin{align} 
\mpjpe(\vec{p}, \widehat{\vec{p}}) = \frac{1}{N} \sum_{i=1}^N \| \vec{p}_i - \widehat{\vec{p}}_i \|_2.
\end{align}
PA-MPJPE, Procrustes aligned MPJPE, is MPJPE calculated after each estimated 3D pose in the batch is aligned to its respective ground truth by the Procrustes method, which is simply a similarity transformation.

All metrics in Table~\ref{table:protorez_vs_baselines} are computed on test sets of AMASS and H36M using models trained on respective training sets using the randomized effector benchmark framework described in detail in~\cite{oreshkin2021protores}.

\end{document}